\documentclass[12pt]{article}

\usepackage[utf8]{inputenc}
\usepackage{amsmath, amsthm, amssymb, color, mathbbol}
\usepackage{graphicx}
\usepackage{makecell}
\usepackage{multicol}

\usepackage[table,xcdraw]{xcolor}
\usepackage{tikz}
\usepackage{array,amscd}
\usepackage{amsfonts}
\usepackage[T1]{fontenc} 
\usepackage{enumerate}
\usepackage{appendix}
\usepackage[arrow, matrix, curve]{xy}

\usepackage{booktabs}
\usepackage{siunitx}
\usepackage{multirow}
\usepackage[font=small,labelfont=bf]{caption}

\usepackage{tikz}
\usetikzlibrary{snakes}
\usetikzlibrary {shapes}
\usetikzlibrary {arrows}
\usetikzlibrary {positioning}
\usetikzlibrary {calc}

\oddsidemargin \evensidemargin
\usepackage{enumitem}
\setenumerate{leftmargin=*}
\allowdisplaybreaks[3]

\makeatletter
\renewcommand*\env@matrix[1][\arraystretch]{%
  \edef\arraystretch{#1}%
  \hskip -\arraycolsep
  \let\@ifnextchar\new@ifnextchar
  \array{*\c@MaxMatrixCols c}}
\makeatother

\usepackage{authblk}
\usepackage{subfig}
\usepackage{threeparttable}
\usepackage{footnote}
\usepackage{booktabs}

\usepackage[utf8]{inputenc}
\usepackage[english]{babel}
 \usepackage{xcolor}
   
\usepackage{amsthm}
\usepackage{afterpage}

\usepackage[linesnumbered,ruled,vlined]{algorithm2e}

\newcommand\blankpage{%
    \null
    \thispagestyle{empty}%
    \addtocounter{page}{-1}%
    \newpage}

\usepackage[utf8]{inputenc} % allow utf-8 input
\usepackage[T1]{fontenc}    % use 8-bit T1 fonts
\usepackage{hyperref}       % hyperlinks
\usepackage{url}            % simple URL typesetting
\usepackage{booktabs}       % professional-quality tables
\usepackage{amsfonts}       % blackboard math symbols
\usepackage{nicefrac}       % compact symbols for 1/2, etc.
\usepackage{microtype}      % microtypography
\usepackage{lipsum}
\usepackage{graphicx}
\usepackage[square,sort,comma,numbers]{natbib}

\title{\textbf{Towards global monitoring: equating the Food Insecurity Experience Scale (FIES) and food insecurity scales in Latin America}}

\author{Federica Onori, Sara Viviani and Pierpaolo Brutti}

\date{}

\begin{document}

\maketitle

\theoremstyle{definition}

\theoremstyle{remark}

 \affil[1]{Department of Statistical Sciences,
   University of Rome La Sapienza,
      Rome}
 %\affil[2]{Department of Statistical Sciences,
 %  University of Rome La Sapienza,
 %  Rome}
 \affil[*]{Corresponding author:  Federica Onori ,   onori.federica@gmail.com}

%\author{} % to eliminate ....

  %% \AND
  %% Coauthor \\
  %% Affiliation \\
  %% Address \\
  %% \texttt{email} \\
  %% \And
  %% Coauthor \\
  %% Affiliation \\
  %% Address \\
  %% \texttt{email} \\
  %% \And
  %% Coauthor \\
  %% Affiliation \\
  %% Address \\
  %% \texttt{email} \\
%}

\maketitle

\begin{abstract}
In order to face food insecurity as a global phenomenon, it is essential to rely on measurement tools that guarantee comparability across countries. Although the official indicators adopted by the United Nations in the context of the Sustainable Development Goals (SDGs) and based on the Food Insecurity Experience Scale (FIES) already embeds cross-country comparability, other experiential scales of food insecurity currently employ national thresholds and issues of comparability thus arise. In this work we address comparability of food insecurity experience-based scales by presenting two different studies. The first one involves the FIES and three national scales (ELCSA, EMSA and EBIA) currently included in national surveys in Guatemala, Ecuador, Mexico and Brazil. The second study concerns the adult and children versions of these national scales. Different methods from the equating practice of the educational testing field are explored: classical and based on the Item Response Theory (IRT).
\end{abstract}

\blankpage{}
% keywords can be removed
%\keywords{Food insecurity,  Item Response Theory (IRT), Test equating, SDGs, Experience-based scales}

\section{Introduction}
\label{sec:1}

Food security is a subject of indisputable relevance, being it conceived as a basic human right since $1948$, as stated in Article $25$ of the Universal Declaration of Human Rights: ``Everyone has the right to a standard of living adequate for the health and well-being of himself and of his family, including food, clothing, housing and medical care'' \citep{united1949universal}. However, food security is a complex and multifaceted concept whose terminology has long been affected by a variety of sectors and disciplines strictly related to it (e.g. agriculture, nutrition, economy, public policy, etc... ) \citep{jones2013we, cafiero2014validity}. A consensus around the definition of food security was finally reached during the World Food Summit in  $1996$ when it was formalized as follows: ``Food security exists when all people, at all times, have physical and economic access to sufficient, safe and nutritious food that meets their dietary needs and food preferences for an active and healthy life" \citep{fao1996world}.\footnote{This definition was further refined in  $2002$ \citep{fao2002state}, when food access was not only conceived in terms of affordability and physical access, but also in terms of removal of \emph{social barriers}. The community of researchers, practitioners and political decision makers currently agree upon the following definition:
  \begin{quote}
Food security exists when all people, at all times, have physical, social and economic access to sufficient, safe and nutritious food that meets their dietary needs and food preferences for an active and healthy life. 
 \end{quote} 
} Grounding on this definition, the conceptualization and operationalization of food security emerge as that of a multidimensional phenomenon made up of four different, hierarchically ordered dimensions: availability, access, utilization and stability \citep{pinstrup2009food, cafiero2013we}. As a consequence, no single indicator can be successfully designated to return a thorough picture of the phenomenon, but a suite of indicators exists, each monitoring specific aspects of food security at different levels of the observation: national, regional, households and individual \citep{jones2013we}. Among all possible aspects related to food insecurity, the dimension of \emph{access to food} is given nowadays high-priority, being acknowledged among the $17$ Sustainable Development Goals (SDGs) of the $2030$ Agenda for Sustainable Development adopted by the United Nations. Access to food is in fact the subject of Target $2.1$ \citep{world2016sustainable}, which states:\\ 
% %\vspace{-.55cm}
 \begin{quote}
 By 2030, end hunger and ensure access by all people, in particular the poor and people in vulnerable situations, including infants, to safe, nutritious and sufficient food all year round.
 \end{quote}
%  %\vspace{-.2cm}
 
 Although food security is now a well-established concept within the scientific community, its definition changed throughout the last century and so did the tools employed to measure the phenomenon \citep{cafiero2014validity, jones2013we}. A brief summary of the main steps will enable to fully appreciate the novelties brought about by the measurement tools developed since the '$80$s. During the $1940$s and for some decades on, the issue of food security was completely identified with that of having enough provisions to cover the needs of the population and, therefore the ``food problem" was mainly dealt with in terms of country-level supplies \citep{fao1946, fao1952, fao1963}. Nevertheless, this formulation could not catch the aspect, yet observable, of malnutrition and famines in countries that did not suffer from food supply at national level \citep{carlson1999measuring}, signal that a change in approaching the measurement of food insecurity was required, in particular towards considering the point of view of people's \emph{access to food}. To mark this change in prospective, the expression \emph{household food insecurity} began to be used.
Since then, other shifts pertained to the definition of food insecurity as for what we use today. A very fundamental one was in the $1990$s when interest moved from dietary energy adequacy to \emph{experience of food insecurity} and livelihood conditions, which involved, among others, also social, nutrition and psychological considerations. Food insecurity has in fact been recognized as a ``managed process", described by means of a spectrum of behaviours and coping strategies that can reveal the level of severity of a food access condition \citep{radimer1992understanding}.
Although specific attitudes and coping strategies might change from country to country, there is a general consensus in the scientific community about the common pattern of behaviours that characterize food insecurity with very minor differences across cultures  \citep{coates2006commonalities}. To this regard, ethnographic and societal studies established that, in case of increasing lack of money or other resources, a common pattern of experiences and behaviours manifests in order to cope with shortage of food \citep{ radimer1992understanding}: at first, psychological concern arises since people start worrying about having enough food; then, a change in the diet occurs by decreasing the quality and variety of the consumed food in order to face a concrete limited access to food; and, in case of more severe food shortages,  people would diminish the quantity of consumed food by reducing meals' size and then by even skipping meals, potentially up to experiencing hunger. The steps just described are commonly referred to as the three \textit{domains} of resource-constrained access to food: psychological concern, decrease of food quality, decrease of food quantity and hunger \footnote{The aim of this first part of the work was mainly to provide a general framework for the topic and clarify that the expression ``food insecurity'' technically refers to a multitude of aspects that relate to different \emph{dimensions}. However, in order to avoid confusion and enable an agile treatise of the subject, hereafter ``food insecurity'' will specifically be meant at the individual or household level and interpreted as the set of the restrictions in accessing food due to limited resources (or, equivalently, \emph{resources-constrained access to food}). This choice will also facilitate conceiving food insecurity as a measurable construct.}.

Mirroring these shifts in the paradigm (from global and national to households and individuals; from food supplies to livelihood conditions; and from objective to subjective measures \citep{cafiero2014validity}), a number of indicators have been proposed to measure food insecurity, like measures of adequacy of food consumption, prevalence of undernourishment, dietary diversity score, etc... Among all, \emph{experience-based food insecurity scales} found a place of relevance, having proved to be a valid and reliable tool for measuring food insecurity in its access dimension, encompassing the current definition of the phenomenon while adopting a behavioural perspective \citep{cafiero2014validity}. As the name suggests, experience-based food insecurity scales measure access to food from a behavioural perspective, building on a set of items that directly ask people about their own personal experiences and behaviours related to the three domains of access to food \citep{jones2013we}. The very first experience-based food insecurity scale was the Household Food Security Survey Module (HFSSM), applied yearly in the United States of America since $1995$ for monitoring purposes \citep{hamilton1997household}. As a matter of fact, the HFSSM  pioneered in this field and several countries in Latin America followed this example by developing their own national scales to be included in national surveys for periodical monitoring. In $2004$, Brazil included the \emph{Brazilian Scale of Food Insecurity} (EBIA) into national Brazilian surveys; Haiti, Guatemala and Ecuador, among other countries, developed the \emph{Latin American and Caribbean Food Security Scale} (Escala Latinoamericana y Caribeña de Seguridad Alimentaria - ELCSA); and in $2008$ Mexico developed its adaptation of the ELCSA, called \emph{Mexican Food Security Scale} (EMSA).
Peculiar to these scales is the availability of two different survey modules, one for households with children and one for households without children and made up of a different number of items.
Finally, beside these country-specific applications of the experience-based approach to measuring food insecurity, in $2013$ the Food and Agriculture Organization of the United Nations (FAO) launched the Voices of the Hungry project (VoH) and developed the Food Insecurity Experience Scale (FIES) conceived as a global adaptation of HFSSM and ELCSA \citep{fao2016meth}. The FIES is based on people's responses to only $8$ dichotomous items and, by means of an ad-hoc methodology that grounds on the Item Response Theory (IRT), and more specifically on the Rasch model, it is the first food insecurity measurement system based on experiences that generates \emph{formally comparable} measures of food insecurity across countries. As such, it is one of the official measurement tool for monitoring progresses toward Target $2.1$ of the SDGs, being the scale used to compute the related Indicator $2.1.1$, (\emph{Prevalence of food insecurity at moderate and severe levels based on FIES}) \citep{fao2016meth, ballard2013food, ballard2014better}.
 
Although the national and regional scales proved to be adequate tools for measuring and monitoring access to food within each country \citep{cientifico2012escala,villagomez2014validez}, the need for a global monitoring, such as that sought in the context of the SDGs, raised the issue of comparing results from applications of different scales in different countries \citep{cafiero2013we}. In fact, despite sharing a common evolution, each national scale uses specific thresholds to measure prevalences of food insecurity for \emph{nominally} the same level of severity. Moreover, comparability issues also arises in the context of each national/regional scale, when considering the adult and the children-referenced versions of the same scale. This work aims at filling this gap by addressing comparability issues in the context of the experience-based food insecurity scales within a statistical framework. Specifically, methods and techniques from the statistical field of the educational testing are applied, with the goal of computing thresholds on the different scales that can be considered as ``equivalent".  Both classical and Item Response Theory (IRT)-based techniques  are employed and two different comparability studies are presented:
\begin{enumerate}
\item Comparison between the FIES and national food insecurity scales in use in some countries in Latin America. Equating analyses are conducted between FIES and ELCSA in Guatemala; FIES and ELCSA in Ecuador; FIES and EMSA in Mexico; and between FIES and EBIA in Brazil.
\item  Comparison between household and children-referenced scales within each national context. This analysis is conducted for ELCSA in Guatemala, ELCSA in Ecuador, EMSA in Mexico and EBIA in Brazil.
\end{enumerate}

Data used for the analysis were collected by the single countries and are available for downloading on the internet at the websites of the Statistical National Institutes of Guatemala, Ecuador, Mexico and Brazil. Implementation of the equating methods was performed on the free software R (http://www.r-project.org) using, among others, the packages \emph{RM.weights} \citep{rmweights}, \emph{equate} \citep{albano2016equate} and \emph{plink} \citep{weeks2010plink}.
The remaining of the paper is organized as follows: Section $2$ describes more in depth the features of the experience-based food insecurity scale; Section $3$ presents the data and Section $4$ is devoted to describe the pillars and the methods of the Test Equating; Section $5$ presents the main results; and Section $6$ concludes with some remarks and possible directions for future works. 
%%\vspace{-.7cm}
\section{Experience-based food insecurity scale}\label{experience}

 As already mentioned, the FIES is strongly based on the ELCSA, which in turn represents a common ancestor for other scales in use in Latin America (EMSA, EBIA, etc...). As a consequence, all these scales largely share the same cognitive content of the items, which constitutes the promising ground on which addressing comparability. Nevertheless, the FIES and the national scales show important differences. First of all, national scales measure food insecurity at the household level, while the FIES produces national measures of food insecurity at the \emph{individual level}. Secondly, national scales have a reference period of $3$ months, while the FIES refers to the $12$ months previous to the interview. Thirdly, and perhaps most importantly, national scales compute prevalences of food insecurity following a \emph{deterministic} methodology based on raw scores (number of affirmative responses) and use discrete thresholds (expressed in terms of raw scores) for computing prevalences of food insecurity at different levels of severity. On the other hand, VoH methodology for the FIES is \emph{probabilistic} in nature in that it fits the Rasch model to the data, models access to food by means of a probabilistic distribution and computes prevalences of food insecurity using thresholds on the continuum latent trait.

%%\vspace{-.5cm}
\subsection{National and Regional Scales of Food Insecurity: ELCSA, EMSA, EBIA}

The survey modules on which ELCSA, EMSA and EBIA are built have strong similarities \citep{cientifico2012escala,villagomez2014validez}. They all account for the three domains of the access dimension of food insecurity discussed in the previous section, aim at measuring food insecurity at the \emph{household} level and all adopt the same reference period of $3$ months previous to the day of the module administration. 
As far as the methodology is concerned, ELCSA, EMSA and EBIA agree on a similar procedure that can be summarized in few steps \citep{cientifico2012escala, villagomez2014validez}:
\begin{enumerate}
\item Computation of a \emph{raw score} for each household: by counting the number of items affirmed by that household. Raw scores represent an \emph{ordinal measure} of food insecurity: the highest the raw score, the more severe the level of food insecurity.
\item Computation of prevalences of food insecurity at three levels of severity: mild, moderate and severe. Prevalences are computed as percentages of households in the sample that scored within a certain range expressed in terms of raw scores and with different thresholds depending on whether children live in the household or not (Table \ref{tab:thres}).
\item Data validation. Homogeneity of the items comprising the scale is assessed by fitting the Rasch model to the data. 
\end{enumerate}

Moreover, each national scale makes use of two different versions of the survey module, distinguishing between households with children (i.e. people under the age of $18$ years) and households without children. The first group of survey modules is usually made up of $6$ to $9$ household-referenced items and, for the sake of simplicity, the scale obtained from this set of items will be referred to, in this work, as the \emph{Adult} scale. The second one integrates the first one by adding from $6$ to $7$ extra children-referenced questions and the scale obtained from this set of items will be referred to as the \emph{Children} scale. The two survey modules thus encompass a different number of items and, from each of them, a scale is built that uses different thresholds to compute prevalences of food insecurity that should be meant to reflect the same level of severity. Prevalences derived from the two scales  are then considered jointly in order to derive national prevalences of food insecurity.

%%\vspace{-.5cm}
 \begin{table}[h]
	\centering
		\begin{tabular}{llll} 
		
	   	%	\hline\noalign{\smallskip}
\midrule

\textbf{Scale} & \textbf{Food insecurity} & \textbf{Households}  & \textbf{Households}  \\
& \textbf{Level}& \textbf{without children} &\textbf{with children}\\ 
%\noalign{\smallskip}\svhline\noalign{\smallskip}

\midrule
\multirow{3}{*}{\textbf{ELCSA}} & mild & 1 to 3 & 1 to 5\\
 & moderate & 4 to 6 & 6 to 10 \\

 & severe & 7 to 8 & 11 to 15\\ \hline
\multirow{3}{*}{\textbf{EMSA}}  & mild & 1 to 2 & 1 to 3 \\
 & moderate & 3 to 4 & 4 to 7 \\

 & severe & 5 to 6 & 8 to 12 \\ \hline

\multirow{3}{*}{\textbf{EBIA}} & mild & 1 to 3 & 1 to 5 \\
 & moderate & 4 to 6 & 6 to 10 \\

 & severe & 7 to 8 & 11 to 15 \\
\bottomrule
		\end{tabular}
	\caption{Classifications of food insecurity using national scales (ELCSA, EMSA and EBIA) and corresponding ranges of the raw scores for households with and without children.
	%and corresponding deterministic thresholds in terms of raw scores.
	}
	\label{tab:thres}
\end{table}

It is worth highlighting that, as reported in Table \ref{tab:thres}, the thresholds used to compute categories of food insecurity that \emph{nominally} reflect the same level of severity (mild, moderate or severe), are country (or regional)-specific. As a matter of fact, these thresholds were not chosen in order to assure comparability among countries (no matter how geographically close to each other they might be) nor in light of clear statistical properties, but according to opinions of experts from the nutrition and social sciences fields. The same consideration holds for the thresholds chosen for the household referenced-scale and the children-referenced scale within each national context.
As a consequence, there is no clear guarantee that, for example, a raw score of $7$ truly reflects the same level of severity in Mexico and Brazil, or that, applying ELCSA in Guatemala, $7$ and $11$ can be considered as equivalent scores in households without and with children, respectively.

%%\vspace{-.4cm}
\subsection{The Food Insecurity Experience Scale (FIES)}

Inspired by Target $2.1$ of the SDGs, the Voices of the Hungry (VoH) project of the Food and Agriculture Organization developed the Food Insecurity Experience Scale (FIES), designed to have cross-cultural equivalence and validity in both developing and developed countries, aiming at producing comparable prevalences of food insecurity at various levels of severity \citep{fao2016meth}. As reported in Table \ref{tab:fiessm}, the FIES Survey Module is made up of $8$ dichotomous items accounting for the three domains of access to food. Since $2014$, the FIES Survey Module (FIES-SM) is part of the Gallup World Poll (GWP) Survey, from Gallup Inc. \citep{tortora2010gallup}, a survey that is repeated every year in over $150$ countries and administered to a sample of adult individuals (aged $15$ or more) representative of the national population. This has practically allowed to reach countries that do not have a national measurement system for food insecurity, yet. In accordance with the characteristics of the GWP,
the version of the FIES-SM here considered refers to a period of $12$ months prior to the survey administration and investigates food insecurity at the level of adult individuals (people aged older than $15$ years), which represents a first difference between FIES and the national scales.

%%%\vspace{-.2cm}
\begin{table}[h]
  \centering
  \begin{tabular}{ll}
%    \toprule
  %  \multicolumn{2}{c}{FIES Survey Module. %English version } \\
\textbf{Items} & \textbf{Abbreviations} \\
%\hline
\midrule
    During the last 12 months, was there a time when,\\ because of lack of money or other resources:\\ \\

    1.~You were worried you would not have enough food to eat? 
 & WORRIED\\[.5\normalbaselineskip]
 
     2.~You were unable to eat healthy and nutritious food?  
 & HEALTY\\[.5\normalbaselineskip]
 
     3.~You ate only a few kinds of foods?  
 & FEWFOOD\\[.5\normalbaselineskip]
 
     4.~You had to skip a meal?  
 & SKIPMEAL\\[.5\normalbaselineskip]
 
     5.~You ate less than you thought you should? 
 & ATELESS\\[.5\normalbaselineskip]
 
      6.~Your household ran out of food?
 & RUNOUT\\[.5\normalbaselineskip]
 
      7.~You were hungry but did not eat?  
 & HUNGRY\\[.5\normalbaselineskip]
 
      8.~You went without eating for a whole day?  
 & WHLDAY\\
 %[.5\normalbaselineskip]
 %   \hline	
 \bottomrule
    \end{tabular}
  \caption{FIES Survey Module (FIES-SM) for individuals and with a reference period of $12$ months.}
  	\label{tab:fiessm}
 
\end{table}

However, the main difference between the two is in the methodology used \citep{fao2016meth}. The Voh methodology developed for the FIES employs a probabilistic model not only as a validation tool (for assessing homogeneity of the items in the scale), but also for computing measurements of food insecurity. In fact, food insecurity is treated as a \emph{latent trait} whose measurement is achieved by means of some ``observables" (the items' answers) and a probabilistic model that links the two. The Rasch model (also known as the one-parameter logistic model or 1PL model) is one of the most simple model that can serve this purpose while, at the same time, assuring a set of favourable measurement properties  \citep{rasch1960probabilistic, fischer2012rasch}. It was proposed in the context of educational testing, where the purpose is generally to score students based on a set of questions (items) and, according to this model, the probability of a respondent to correctly answering the $j-$th item is modelled as a logistic function of the distance between two parameters, one representing the item's severity ($b_{j}$) and one representing the respondent's ability ($\theta$):

\begin{equation}
P_{j}(\theta) = P(X_{j} = 1|\theta;b_{j}) = \frac{\exp(\theta - b_{j})}{1 + \exp(\theta - b_{j})}.
\end{equation}

\begin{figure}[h]
\centering
\includegraphics[width=12 cm,height=9 cm,keepaspectratio]{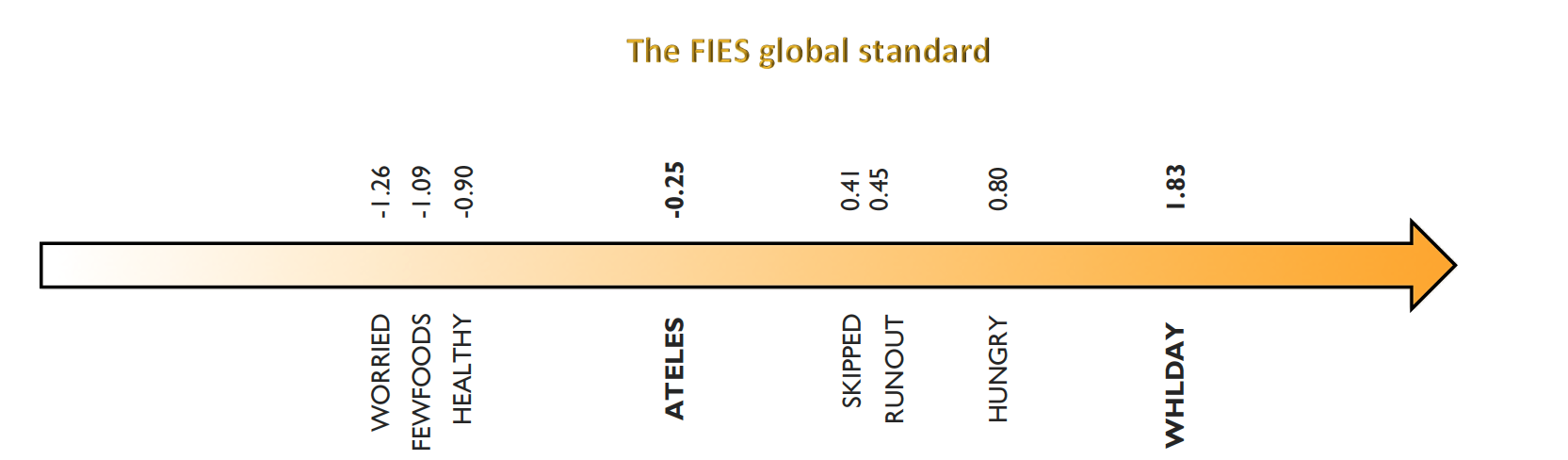}
\caption{The FIES Global Standard.}
\label{fig:fiesgs}
\end{figure}

\begin{figure}[h]
\centering
\includegraphics[width=12 cm,height=8 cm,keepaspectratio]{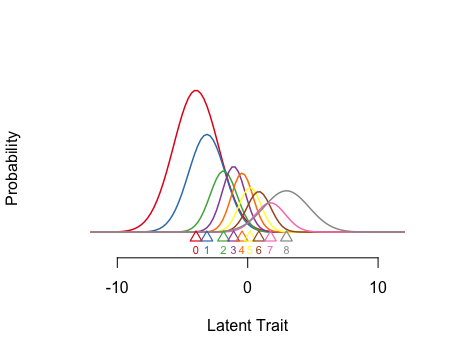}
\caption{Distributions of  severity of food insecurity among respondents according to their raw scores}
\label{fig:passprob}
\end{figure}

The Rasch model provides a sound statistical framework to assess the suitability of a set of items for scale construction and comparing performance of scales. Basic assumptions are unidimensionality, local independence, monotonicity, equal discriminating power of the items and logistic shape of the Item Response Functions (IRFs). Moreover, it has several interesting properties for which it earned its success among social science measurement models, like sufficiency of the raw score, independence between items and examinees' parameters, and invariance property \citep{hambleton1991fundamentals}. In the context of food insecurity, the item's severity can be interpreted as the severity of the restrictions in food access represented by each item while the ability parameters are to be meant as the overall severity of the restrictions in accessing food that the respondent had to face (in light of her answers to the items in the survey module).
From the point of view of the Stevenson's classification of scales \citep{stevens1946theory}, this way of measuring food insecurity guarantees the construction of an \emph{interval} scale as opposed to the \emph{ordinal} scale obtained from the methodology employed, for instance, for ELCSA, EMSA and EBIA and named \emph{deterministic} as opposed to the \emph{probabilistic} developed for the FIES. Moreover, prevalences obtained by means of the FIES are guaranteed to be comparable across countries, thanks to the  implementation of an \emph{equating step} for which estimates of the model parameters obtained in a single application of the scale are adjusted on the FIES Global Standard scale, a set of item parameters serving as a reference metric and based on application of the FIES in all countries that were covered by the GWP survey in $2014$, $2015$ and $2016$ \citep{fao2016meth} (Fig. \ref{fig:fiesgs}). Finally, each respondent is assigned a probabilistic distribution of his/her food insecurity along the latent trait, depending on his/her raw score. This distribution is Gaussian with mean equal to the adjusted (to the Global Standard) respondent parameter and standard deviation
equal to the adjusted measurement error for
that raw score. As a last step, this mixture of distributions is used to compute the percentage of population whose severity is beyond a fixed threshold on the latent trait, calculated as a weighted sum across raw scores, with weights reflecting the proportions of raw scores in the sample (Figure \ref{fig:passprob}). While theoretically it is possible to compute percentages of population beyond each and every value on the continuum, the VoH methodology suggests the computation of two prevalence rates corresponding to choosing thresholds on the Global Standard metric set at the severity of items ATELESS ($-0.25$) and WHLDAY ($1.83$) (Fig. \ref{fig:fiesgs}). %This choice is motivated by the need of measuring the phenomenon at different levels of severity and 
The resulted indicators of food insecurity take the name of 
\emph{Prevalence of Experienced Food Insecurity at moderate or severe levels} ($FI_{Mod+Sev}$) and \emph{Prevalence of Experienced Food Insecurity at severe levels} ($FI_{Sev}$), respectively.
However, in order for these quantities to be valid and reliable measurements of food insecurity, a validation step must be undertaken in each and every application of the scale. This is commonly performed by computing goodness-of-fit statistics of the Rasch model (e.g. Infit, Outfit and Rasch reliability statistics) that assess the good behaviour of the items and by performing a Principal Component Analysis (PCA) on the residuals to investigate the existence of a second latent trait. For more details on the usage of the Rasch model as a measuring tool for food insecurity we refer the reader to \citep{marknord}, while for more insights in the VoH methodology for the FIES we refer to \citep{fao2016meth}.


%%\vspace{-.6cm}
\newpage

\section{Data}
Data referred to Guatemala were collected in the \emph{Encuesta Nacional de Condiciones de Vida} (ENCOVI) conducted
by the \emph{Instituto Nacional de Estadística} (INE) in $2014$ and the sample used included $11433$ households. Data referred to Ecuador were collected in the \emph{Encuesta Nacional de Empleo y Desempleo} (ENCOVI) conducted
by the \emph{Instituto Nacional de Estadísticas y Censos} (INEC) in $2016$ and the sample used included $16716$ households. Data referring to Mexico were collected in the \emph{Encuesta Nacional de Ingresos y Gastos de los Hogares}  (ENIGH) conducted
by the \emph{Instituto Nacional de Geografia e Estatística} in $2014$ and the sample used included $19479 $ households. Data referring to Brazil were collected in the \emph{Pesquisa Nacional de Amostra de Domicílios} (PNAD) conducted
by the \emph{Instituto Nacional de Geografia e Estatística} (IBGE) in $2013$ and the sample used included $116543$ households. 
All samples were representative of the corresponding national populations.

\label{sec:2}
%%\vspace{-.5cm}
%\subsection{Item response Models}
\section{Test Equating}

Scores deriving from tests usually have an important role in the decision making process that brings to excluding some candidates for a job or scholarship position, or adopting specific public policy strategies in order to take action on a public relevant issue. 
Evidently, this requires that tests to be administered in multiple occasions, as it is the case for the admission college tests that are held in specific \emph{test dates} during the year. Therefore, a crucial consideration arises: if the same questions were included in the tests, students that already took the test would have an advantage and the test would rather measure the degree of exposure of students to past tests than their ability on some specific subject. At the same time, it is important that all students take the ``same" test, in order to fairly compare performances and make decisions accordingly. This issue is commonly addressed by administering on every test date a different version of the same test, called \emph{test form}, that is built according to certain content and statistical \emph{test specifications}. Nonetheless, minor differences might still occur among different test forms, one resulting slightly more difficult than the others. Therefore, in order to evenly score students that took multiple test forms and establish if a poorer performance is due to a less skillful respondent and not to a more difficult test, a procedure is needed to make tests comparable. This procedure is called \emph{equating} and it is formally defined as the statistical process that is used to adjust for differences in difficulty between tests forms built to be similar in content and difficulty, so that scores can be used interchangeably \citep{kolen2014test, dorans2007linking}. Every test equating should meet some fundamental equating requirements and needs the specification of both a data collection design and of one or more methods to estimate an equating function. All these aspects will be discussed in the remaining of this section. 

%%%\vspace{-.3cm}
\subsection{Equating Requirements} \label{requir}
Equating scores on two test forms $X$ and $Y$ must meet some requirements that assure that the equating to be meaningful and useful (i.e. equated scores can be used interchangeably). The following five requirements are globally considered of primary importance for an equating to be run, although they would better be considered as general guidelines than easily verifiable conditions:
\begin{itemize}
\item \textbf{Equal Construct Requirement}
Tests that measure different constructs should not be equated.
\item \textbf{Equal Reliability Requirement}
Tests that measure the same construct but differ in reliability should not be equated.
\item \textbf{Symmetry Requirement}
Equating function that equate scores on $X$ to scores on $Y$ should be the inverse of the equating function that equate scores on $Y$ to scores on $X$.
\item \textbf{Equity Requirement}
For the examinee should be a matter of indifference which test will be used.
\item \textbf{Population Invariance Requirement}
Equating function used to equate scores on $X$ and scores on $Y$ should be \emph{population invariant} in that the choice of a specific sub-population used to compute the equating function should not matter.
\end{itemize}

It might be the case that the two tests to be equated do not satisfy all five requirements. For example, they could differ in length and statistical specifications, with consequences on the ``Equal Reliability requirement'' and ``Equity requirement''. In fact, a longer test would be in general more reliable and, if a poorly skillful examinee had to be scored, he or she would have more chance to score higher if administered the shortest test. 
The aforementioned requirements assure that scores derived from tests that do meet all of them can be used interchangeably while, if they do not all strictly hold, the exercise would rather be addressed as a weaker analysis of comparability named \emph{linking} \citep{kolen2014test,dorans2007linking}.

%%\vspace{-.3cm}

\subsection{Equating designs}
There are basically two ways in which data collection designs can account for differences in the difficulty of two or more test forms in test equating, namely either by the use of ``common examinees" or the use of ``common items" \citep{kolen2014test,dorans2007linking}. In the first case the same group of examinees  (or two random samples of examinees from the same target population) take both tests. In this case, any difference in the scores is attributable to differences in the test forms. Examples of this category are the ``Single-Group" (SG) and the ``Equivalent-Groups" (EG) designs. In the second case, a set $A$ of common items called \emph{anchor test} is included in both test forms in order to account for such differences. Therefore, any difference between scores on the anchor test is due to differences among examinees. Data designs that use this method are called ``Non-Equivalent groups with Anchor Test" (NEAT) designs.
%%\vspace{-.3cm}
\subsection{Equating Methods}
Several equating methods have been proposed and applied to equate observed scores on equatable tests.
In this section an overview of the most common and popular methods is provided, starting from the observed-score methods of mean, linear and equipercentile equating and ending up with the true score equating in the context of IRT. All these methods have been implemented for the two comparability studies between experience-based food insecurity scales that are presented in this work.

\subsubsection{Observed-score equating methods}
Let $X$ and $Y$ be two tests (or two forms of the same test) scored correct/incorrect (1/0). 
Scores on test $X$ and $Y$ will be denoted as random variables \textbf{X} and \textbf{Y} with possible values, respectively $x_{k}$, ($k=0, \dots, K)$ and $y_{l}$  (for $l=0, \dots, L$), where $K$ and $L$ are the lengths of tests $X$ and $Y$, respectively.
We denote the score probabilities of \textbf{X} and \textbf{Y} by

\begin{equation}
r_{k} = P(\textbf{X} = \textbf{x}_{k}) \quad \textrm{and} \quad s_{l} = P(\textbf{Y} = \textbf{y}_{l}).
\end{equation}
The cdfs of \textbf{X} and \textbf{Y} are denoted by
\begin{equation}
F(x) = P(\textbf{X}\leq x) \quad \textrm{and} \quad G(x) = P(\textbf{Y}\leq y)
\end{equation}
and the moments are, respectively
\begin{equation}
\mu_{X} = \textbf{E}(\textbf{X}), \qquad \mu_{Y} = \textbf{E}(\textbf{Y}) 
\end{equation}
and
\begin{equation}
\sigma_{X} = SD(\textbf{X}), \qquad \sigma_{Y} = SD(\textbf{Y})
\end{equation}

%%\vspace{-.2cm}
\paragraph{Mean equating}
In mean equating, test form $X$ is assumed to differ from test form $Y$ by a constant amount along the scale.
For example, if form $X$ is 2 points easier than form $Y$ for low-ranking examinees, the same will hold for high-ranking examinees.
In mean equating, two scores on different forms are considered equivalent (and set equal) if they are the same (signed) distance from their respective means, that is
\begin{equation}
x - \mu_{X} = y - \mu_{Y}.
\end{equation}
Then, solving for $y$, the score on test $Y$ that is equivalent to a score $x$ on test $X$, and called $m_{Y}(x)$, is
\begin{equation}
m_{Y}(x) = y = x - \mu_{X} + \mu_{Y}.
\end{equation}
Clearly, mean equating allows for the means to differ in the two test forms.

%%\vspace{-.2cm}
\paragraph{Linear equating}
In linear equating, difference in difficulty between the two tests is not constraint to remain constant but can vary along the score scale. 
In this equating method, scores are considered equivalent and  set equal if they are an equal (signed) distance from their means in standard deviation units, that is the two standardized deviation scores (z-scores) on the two forms are set equal
\begin{equation}
\frac{x - \mu_{X}}{\sigma_{X}} = \frac{y - \mu_{Y}}{\sigma_{Y}}
\end{equation}
from which the score on test $Y$ equivalent to a score $x$ on test $X$, and that is called $l_{Y}(x)$, is
\begin{equation}
l_{Y}(x) = y = \frac{\sigma_{Y} }{\sigma_{X} } x +  \left[ \mu_{Y} - \frac{\sigma_{Y}}{\sigma_{X}} \mu_{X}\right]. 
\end{equation}
where $\frac{\sigma_{Y} }{\sigma_{X} }$ can be recognized as the \emph{slope} and $ \mu_{Y} - \frac{\sigma_{Y}}{\sigma_{X}} \mu_{X}$ as the the \emph{intercept} of the linear equating transformation. Linear equating allows for both means and scale units to differ in the two test forms.

%%\vspace{-.2cm}
\paragraph{Equipercentile equating}
In the equipercentile equating method a curve is used to describe differences between scores in the two forms. 
Basic criterion for the equipercentile equating transformation is that the distribution of the scores on Form $X$ converted to the Form $Y$ scale is equal to the distribution of scores on Form $Y$. Scores on the two forms are considered -and set- equivalent if they have the same \emph{percentile rank}. We adopt here the definition of equipercentile equating function given by Braun and Holland in \citep{braun81}. Let's consider the random variables $\textbf{X}$ and $\textbf{Y}$ representing the scores on forms $X$ and $Y$ and $F$ and $G$ their cumulative distribution functions. We call $e_{Y}$ the symmetric equating function converting Form $X$ scores into scores on Form $Y$ scale and $G^{\star}$ the cumulative distribution function of $e_{Y}(\textbf{X})$, that is the cdf of the scores on Form $X$ converted to the Form $Y$ scale. Function $e_{Y}$ is the \emph{equipercentile equating function} if $G^{\star} = G$.
According to the definition of Braun and Holland, if $\textbf{X}$ and $\textbf{Y}$ are continuous random variables, then 
\begin{equation}
e_{Y}(x) = G^{-1}[F(x)],
\end{equation}
is an equipercentile equating function, where $G^{-1}$ is the inverse of $G$.
This definition meets the ``Symmetric requirement'' and, given a Form $X$ score, its equivalent on the Form $Y$ scale is defined as the score having the same percentage of examinees at or below it.

%\vspace{-.5cm}

\subsubsection{IRT-based equating methods: the IRT-True Score equating (IRT-TS)}
\begin{figure}[h]
% \begin{subfigure}
\includegraphics[width=.5\linewidth, height=7.5cm]{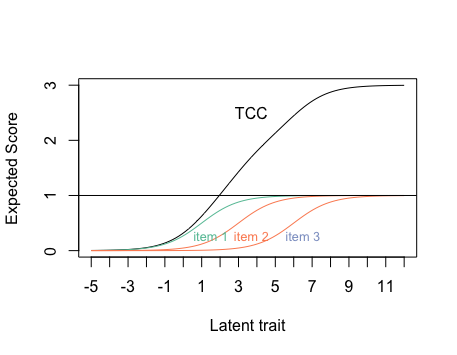} 
%\end{subfigure}
%\begin{subfigure}
\includegraphics[width=0.5\linewidth, height=7.5cm]{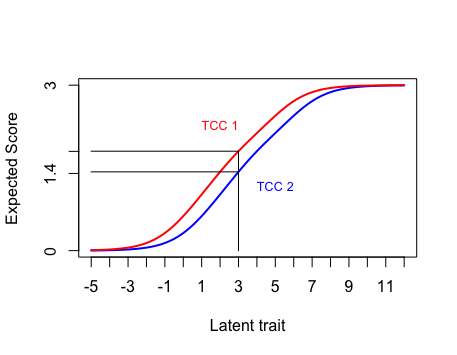}
%\end{subfigure}
 \caption{Test Characteristic Curve (TCC) referred to a test of three items (left) and a pictorial description of the IRT-True Score (IRT-TS) equating method (right).}
\label{fig:IRTTS}
\end{figure}

Equating different forms of the same test using the IRT-True Score equating (IRT-TS) is a three steps process \citep{cook1991irt}:
\begin{enumerate}
\item \emph{Estimation}: Fit an IRT-model to the data for both tests. 

This step consists in assessing goodness-of-fit of a specific IRT model and estimating item parameters for both forms. In the case of the Rasch model, in light of the sufficient statistics property, estimates of the item severities do not depend on the group of examinees and therefore the IRT-TS based on the Rasch model can be claimed to meet the ``Population Invariance requirement", since it produces results that are sample-independent.

\item \emph{Linking}: Put parameters' estimate on a common metric through a linear transformation based on a set $A$ of common items.

In this second step, a linear transformation is used to bring parameter estimates to a common IRT scale. In fact, "if an IRT model fits the data, then any linear transformation of the $\theta$-scale also fits the data, provided that the item parameters are transformed as well" \citep{kolen2014test}. Let consider Form $X$ made up of $J$ dichotomously scored items administered to $N$ examinees and let consider the Rasch model to fit the data.
Then, if $P$ and $Q$ are Rasch scales that differ by a linear transformation, the item severities $b_{j}, j \in\{1, \dots, J\}$  are related as follows
$$
b_{j_{Q}} = A b_{j_{P}} + B,  \qquad j \in\{1, \dots, J\}
$$
and the same relationship holds for the ability parameters. A useful way to express the constants $A$ and $B$ is through the mean and standard deviation of the item parameters in both scales
$$
A = \frac{\sigma(b_{Q})}{\sigma(b_{P})}, \qquad
B = \mu(b_{Q}) - A \mu(b_{P}).
$$

In equating two different forms of the same test with a set of common items administered to non-equivalent groups, it is possible to exploit this linear relationship through the so called \emph{Mean/Sigma transformation method} \citep{marco1977item}, which uses means and standard deviations of item parameter's estimates of only those items in the anchor test. More specifically, given Form $X$ and Form $Y$ with a set $A$ of common items, estimates of the difficulty parameters for items in the set $A$ in the two calibrations are linked via a linear transformation and used to compute the coefficients $A$ and $B$ of this transformation. Once the transformation is estimated, it can be applied to transform the ability parameters on one Form to the corresponding parameters on the other Form, thus enabling comparability between the two Forms.

\item \emph{Equating}: Get equivalent expected raw scores through the Test Characteristic Curves of the two tests (TCC).

Once the metrics of the two Forms are put on the same scale (that can be either the scale of one of them or a third scale) it is finally possible to compare performance of examinees taking the two Forms. However, as it often happens with standardized tests, reported scores could be expressed in terms of raw scores and, if this is the case, a further step is needed. 
Within the framework of IRT, it is possible to mathematically relate ability estimates to specific \emph{true scores} on each test form. The \emph{IRT- True Score equating} method computes equivalent true scores in the two forms and considers them, as it is common in the practice of equating, as equivalent observed scores \citep{lord1983comparison}. 
Given Form $X$ and Form $Y$ two test forms measuring the same ability $\theta$ with respectively $n_{X}$ and $n_{Y}$ items and given both item and ability estimates are on the same scale through a linear transformation, the estimated true scores on the two forms are related to $\theta$ by the so called \emph{Test Characteristic Curve} as follows
$$
T_{X} = \sum_{j=1}^{n_{X}}\hat{P}_{j}(\theta), \qquad
T_{Y} = \sum_{i=1}^{n_{Y}}\hat{P}_{i}(\theta)
$$
where $T_{X}$ (respectively $T_{Y}$) is the estimated true score for Test $X$ ($Y$) and $\hat{P}_{i}(\theta)$ ($\hat{P}_{j}(\theta)$) is the estimated probability function for item $j$ ($i$) (Fig. \ref{fig:IRTTS}).
Through the Test Characteristic Curve, an ability $\theta$ can thus be transformed into an estimated true score on the test form and, provided ability parameters on the two forms are put on the same metric, true scores corresponding to the same $\theta$ are considered equivalent.

\end{enumerate}

%\vspace{-1.5cm}

\section{Results}
As a preliminary step to both equating studies, the Rasch model has been fitted to all eight datasets (an Adult scale and a Children scale in the four countries), and a validation step was performed to confirm the good behaviour of the scale. 
In all eight applications it was possible to observe an overall good fit of the model. Assumptions of equal discrimination of the items was certainly met, thanks to item Infit statistics entirely in the range of $(0.7,1.3)$, confirming the strength and consistency of the association of each item with the underlying latent trait (compare \citep{marknord, fao2016meth}). Moreover, Outfit statistics were never as high as to warn misbehaviour due to highly unexpected response patterns, assessing the good performance of the items. Assumptions of conditional independence and unidimensionality of the items were assessed through computation of conditional correlations among each pair of items and submission of the correlation matrix to principal component factor analysis (PCA). All pairwise residual correlation were, in absolute value, smaller than $0.4$ thus confirming that all correlations among items result from their common association with the latent trait. PCA performed on the matrix of residual correlations  showed the presence of only one main dimension that, due to the cognitive content of the items, can thus be recognized as the \emph{food access} dimension that the scales aim at measuring.  Finally, overall model fit is assessed by Rasch reliability statistics (proportion of total variation in true severity in the sample that is accounted for by the model), ranging between $0.65$ (Mexico) and $0.79$ (Guatemala) for the Adult scale and between $0.80$ (Mexico) and $0.86$ (Guatemala) for the Children scale, confirming a good overall discriminatory power for all scales. Sporadic departures from this irreproachable behaviour could only be attested for one or two items in the Children scale (like a residual correlation of $0.6$ between two item of the ELCSA in Guatemala) that however never compromised the good performance of the overall scale. 
%\vspace{-.7cm}

\subsection{First Study: Equating FIES and National Scales} %

The aim of this comparability study is to find raw scores on the national scales EBIA, EMSA and ELCSA that can be considered \emph{equivalent} to the continuum FIES global thresholds used to compute the two indicators $FI_{Mod+Sev}$ and $FI_{Sev}$, namely $-0.25$ and $1.83$. However, it is worth noticing that, since VoH methodology uses thresholds on the continuum while national scales methodology uses discrete thresholds, the equivalent raw score will almost never exactly produce the same prevalence obtained with the VoH thresholds. 

As it is currently set up and implemented, the FIES Module refers to adults (people aged $15$ or above). Therefore, in order to meet the ``Equal Construct requirement'', the modules of the national scales administered to households \emph{without} children have here been considered. Technically, the FIES Survey Module and the survey modules of the national scales (households without children) will thus serve the role of \emph{test forms} of the same test that are to be equated. This was ultimately made possible in light of the common history that brought to the development of these scales (i.e. FIES, ELCSA, EMSA, EBIA), which assures that, despite some differences such as the level of the measurement and the reference time (see Section \ref{experience}), the survey modules used to collect data have very strong similarities and share the same dichotomous structure (possible answers are ``Yes/No'').

This first study was carried out by implementing the following methods:

\begin{enumerate}
    \item \textbf{IRT True Score} (IRT-TS) equating. 
    \item \textbf{Linking} via a linear transformation applied to ability parameters.
    \item \textbf{Minimization} of the difference between prevalences of food insecurity.
\end{enumerate}

The IRT-TS equating method was implemented in the context of the  NEAT equating design. In this work, the set $A$ of common items was computed according to an iterative procedure that starts with all items considered as in common (apart from the ones classified as unique \emph{a priori}) and then discards one item at a time beginning from the one that exceeds the tolerance threshold of $0.5$ the most. Algorithm ends when a set $A$ of items all within this threshold is found.  Item WHLDAY was considered as \emph{unique} a priori in all four equating analyses due to its different cognitive content in the considered scales: more severe in the FIES since it refers to ``not eating for a whole day'', and less severe in the national scales where it reports on members of the household that either only ate once \emph{or} went without eating for a whole day. The Standard Error of Equating (SEE) for the IRT True-Score equating method was estimated using $1000$ bootstrap
replications \citep{kolen2014test}. The second and third methods can be considered as either variations of the IRT-TS or techniques that might sound particularly reasonable in the present context. They were explored for investigation purposes and the obtained scores won't be claimed to be ``equivalent", but rather ``corresponding" scores. In fact, the second method (Linking) consists in considering the linear transformation obtained in the second step of the IRT-TS method and applying it to the estimated ability parameters of the Rasch model. Once ability parameters are adjusted to the Global Standard metric, the raw score corresponding to the ability parameters that are closer to the two VoH thresholds are considered as \emph{corresponding} raw score. On the other hand, the third method (Minimizing) consists in computing prevalences of food insecurity at the household level applying the FIES methodology to the data used for the national scales and comparing the prevalences so obtained with the percentages of population scoring from a certain raw score on. The two raw scores that realize the minimum distance with the two VoH global thresholds (in terms of prevalences) are considered as the \emph{corresponding} raw scores in accordance to this method. 

Results from the first comparability study are summarized in Table \ref{tab:modsev} and Table \ref{tab:sev}, which report the raw scores on the national scales that are computed equivalent to the VoH global thresholds used for the indicators $FI_{Mod+Sev}$ and $FI_{Sev}$, respectively. Table \ref{tab:modsev} shows that the threshold used for computing $FI_{Mod+Sev}$ and corresponding to the severity of item ATELESS on the Global Standard metric (i.e. $-0.25$) might reflect a \emph{less severe} condition of food insecurity compared to the one measured by the national scales for the moderate category of food insecurity. In fact, all the equated raw scores are either equal to or around one point less than the thresholds currently used by ELCSA, EMSA and EBIA for this category of food insecurity. On the contrary, the threshold used for $FI_{Sev}$ and corresponding to the severity of item WHLDAY on the Global Standard metric (i.e. $1.83$) generally reflects a \emph{more severe} condition of food insecurity than the one captured by the national scales for the severe level of food insecurity, Table \ref{tab:sev} reporting equated raw scores that are either equal to or one point higher than the national thresholds currently in use for this category.
%\vspace{-.4cm}

  \begin{table}[h]
	\centering
		
		\begin{tabular}{lllll}

	%	\hline\noalign{\smallskip}
\toprule
\textbf{Food Insecurity} & \textbf{Internal} & \textbf{IRT-TS} &  \textbf{Linking} & \textbf{Min. Diff.}\\\textbf{Scales} &\textbf{Monitoring} & \textbf{Rasch (SEE)}&  &\\
%\noalign{\smallskip}\svhline\noalign{\smallskip}
\midrule
ELCSA (Guatemala)&	4&	3.3 (0.19)&	3&	4\\
ELCSA (Ecuador)&	4&	4.2 (0.14)&	4&	4\\
EMSA (Mexico)&	3&	2.0 (0.23)&	2&	2\\
EBIA (Brazil)&	4&	4.0 (0.09)&	4&	5\\
\bottomrule
%\hline
		\end{tabular}
	\caption{Equated Raw Scores on the national scales corresponding to the VoH threshold for $FI_{Mod+Sev}$ ($-0.25$ on the Global Standard).}
	\label{tab:modsev}
\end{table}

%\vspace{-1.2cm}

 \begin{table}[h]
	\centering
		\begin{tabular}{lllll}

\toprule
%		\hline\noalign{\smallskip}

\textbf{Food Insecurity} & \textbf{Internal} & \textbf{IRT-TS} & \textbf{Linking} & \textbf{Min. Diff.}\\\textbf{Scales} &\textbf{Monitoring} & \textbf{Rasch (SEE)}& &\\
%\noalign{\smallskip}\svhline\noalign{\smallskip}
\midrule
ELCSA (Guatemala)&	7&	7.8 (0.18)&		8&	8\\
ELCSA (Ecuador)&	7&	7.1 (0.18)&		7&	8\\
EMSA (Mexico)&	5&	6.0 (0.26)&		6&	6\\
EBIA (Brazil)&	6&	7.9 (0.07)&		8&	8\\
\bottomrule
		\end{tabular}
		\\
		
	\caption{Equated Raw Scores on the national scales corresponding to the VoH threshold for $FI_{Sev}$ ($1.83$ on the Global Standard).}
	\label{tab:sev}
\end{table}

%\vspace{-1.1cm}

\subsection{Second Study: Comparing Household- and Children-referenced item scales} %

This second analysis aims at comparing the Adult and Children scales within each national context. To this purpose, we implemented the Single Group (SG) data collection design by considering the scores obtained by the households with children on both survey modules (the one containing only adult and household-referenced questions and the one containing also children-referenced questions). This equating design is usually not easily implemented, since it requires the same group of respondents to be administered two different forms of the same test, resulting in an expensive and time-consuming procedure. However, we here could exploit the fact that the survey module for the Children scale is simply an extended version of the module for the Adult scale (see Section \ref{experience}) and as such, we can ``imagine'' to administer the adult referenced survey to households with children just by dropping children-referenced items. With regards to the equating requirements (see Section \ref{requir}), it is worth noticing that the two survey modules have different length and, as such, the obtained scales could have different reliability which, in turn,  could potentially challenge the ``Equal Reliability Requirement''.  
Equating of the Adult and Children scales in the four countries was carried out through implementation of four equating methods: IRT True Score equating with the Rasch model, Mean, Linear and Equipercentile  equating methods \citep{kolen2014test}. The first method is IRT-based while the other four are classical methods of equating, that do not rely on any model to fit the data but only on the observed raw scores.

Tables \ref{tab:equatemethE} and \ref{tab:equatemethB}  show raw scores on the Children scale that are computed equivalent to raw scores on the Adult scales that are used as \emph{lower} thresholds for the moderate and severe categories of food insecurity. Results suggest that the national thresholds currently used for nominally the same levels of severity could reflect different degrees of the severity of access to food. This is particularly evident when looking at the most severe category of food insecurity, where raw scores on the Children scale that are computed equivalent to the lower thresholds on the Adult scale are around one point \emph{higher} than the thresholds currently in use for households with children for ELCSA in Guatemala and EMSA in Mexico and between one and two points \emph{lower} for EBIA (Tables \ref{tab:equatemethE} and \ref{tab:equatemethB}, column ``Severe"). On the other hand, the corresponding raw scores for moderate food insecurity mainly align with the thresholds currently in use for this category (Tables \ref{tab:equatemethE} and \ref{tab:equatemethB}, column ``Moderate").
Interestingly, minor differences emerge between the behaviour of the equated scores for ELCSA in Guatemala and Ecuador, possibly due to specific features of the phenomenon in the two countries, confirming the importance of an equating analysis even between different applications of the same scale. Finally, it is noteworthy that, among all implemented methods, the Equipercentile equating method is the one whose results mostly resemble the current adopted thresholds.

%\vspace{-.3cm}

\begin{table}[h]
	\centering
	\begin{tabular}{lr} 
	\centering

	\begin{tabular}{lll} 
			%	\hline\noalign{\smallskip}
\toprule
\textbf{Equating}& \textbf{Moderate}& \textbf{Severe}\\
\textbf{Method} &  \textbf{(SEE)}& \textbf{(SEE)} \\
%\noalign{\smallskip}\svhline\noalign{\smallskip}

\midrule
	IRT-TS&	6.2 (0.09)	&	12.1 (0.10)	\\
	Mean&	6.6	(0.07)&	12.2	(0.07)\\
	Linear&	6.5	(0.07)&	11.7	(0.11)\\
	Equip&	6.3	(0.09)&	11.3	(0.15)\\
	%\hline
%	Kernel Equip&	6.1&	(0.02)&	12.0&	(0.04)\\
\bottomrule

		\end{tabular} 
\qquad \qquad

			\begin{tabular}{lll} 
			%	\hline\noalign{\smallskip}
\toprule
		\textbf{Equating}& \textbf{Moderate}& \textbf{Severe}\\
\textbf{Method} &  \textbf{(SEE)}& \textbf{(SEE)} \\
%\noalign{\smallskip}\svhline\noalign{\smallskip}
\midrule
IRT-TS&	5.8 (0.09)	&	12.1 (0.11)	\\	
Mean&	6.4	(0.05)&	11.6	(0.05)\\	
Linear&	6.2	(0.07)&	11.0	(0.10)\\	
Equip&	5.8	(0.13)&	11.2	(0.16)\\	
%\hline
%Kernel Equip&	5.4&	0.02&	11.9&	0.04\\	

		\bottomrule
\end{tabular}

\end{tabular}

\caption{Raw scores on the Children scale corresponding to raw scores $4$ and $7$ on the Adult scale (lower thresholds for moderate and severe food insecurity, respectively) and related Standard Error of Equating (SEE) computed by means of the IRT-TS, Mean, Linear and Equipercentile equating methods. Left: ELCSA (Guatemala). Right: ELCSA (Ecuador)}
	\label{tab:equatemethE}
\end{table}

%\vspace{-.5cm}

\begin{table}[h]
	\centering
	\begin{tabular}{lr}  
	\centering

	\begin{tabular}{lll} 
			%	\hline\noalign{\smallskip}
\toprule
\textbf{Equating}& \textbf{Moderate}& \textbf{Severe}\\
\textbf{Method} &  \textbf{(SEE)}& \textbf{(SEE)} \\
%\noalign{\smallskip}\svhline\noalign{\smallskip}
\midrule

IRT-TS&	4.8 (0.12)	&	8.7 (0.13)	\\
Mean&	5.5	(0.05)&	9.5	(0.05)\\
Linear&	5.1	(0.07)&	8.6	(0.10)\\
Equip&	4.8	(0.13)&	8.1	(0.14)\\
%\hline
%Kernel Equip&	4.5&	(0.03)&	8.8& (0.04)\\	
\bottomrule
		\end{tabular} 
\qquad \qquad

			\begin{tabular}{lll} 
	%			\hline\noalign{\smallskip}
\toprule
		\textbf{Equating}& \textbf{Moderate}& \textbf{Severe}\\
\textbf{Method} &  \textbf{(SEE)}& \textbf{(SEE)} \\
%\noalign{\smallskip}\svhline\noalign{\smallskip}

\midrule

IRT-TS&	4.8 (0.08)	&	8.7 (0.09)	\\
Mean&	5.5	(0.04)&	9.0	(0.04)\\
Linear&	5.5	(0.04)&	8.8	(0.07)\\
Equip&	4.7	(0.05)&	8.3	(0.14)\\
%\hline
%Kernel Equip&	5.4&	(0.02)&	9.3&	(0.04)\\
		\bottomrule
\end{tabular}

\end{tabular}
	\caption{Left: Raw scores on the Children scale corresponding to raw scores $3$ and $5$ on the Adult scale (lower thresholds for moderate and severe food insecurity, respectively) and related Standard Error of Equating (SEE) computed by means of the IRT-TS, Mean, Linear and Equipercentile equating methods, EMSA (Mexico). Right: Raw scores on the Children scale corresponding to raw scores  $4$ and $7$ on the Adult scale (lower thresholds for moderate and severe food insecurity, respectively) and related Standard Error of Equating (SEE) computed by means of the IRT-TS, Mean, Linear and Equipercentile equating methods, EBIA (Brazil).}
	\label{tab:equatemethB}
\end{table}

%%\vspace{-.5cm}

\section{Conclusions}
The present work presented two studies investigating comparability between experiential scales of food insecurity.
    The first study aimed at addressing comparability between the FIES and three national scales (ELCSA, EMSA and EBIA) in Guatemala, Ecuador, Mexico and Brazil. Results show that, in general, the VoH threshold used for computing the indicator \emph{Prevalence of Experienced Food Insecurity at moderate or severe levels} ($FI_{Mod+Sev}$)  and corresponding to the severity of item ATELESS on the Global Standard ($-0.25$) seems to reflect a \emph{less severe} level of food insecurity than that described by the thresholds used by the national scales (and expressed in terms of raw scores) for the same level of severity. On the other hand, the VoH threshold used for computing the indicator \emph{Prevalence of Experienced Food Insecurity at severe levels} ($FI_{Sev}$) and corresponding to the severity of item WHLDAY on the Global Standard ($1.83$) seems to reflect a \emph{more severe} condition than the one measured through the national thresholds for this level of severity.
The relevance of such a result for the practice of food insecurity measurement is self-evident. In fact, the possibility of comparing prevalences of food insecurity derived from applying different scales represents an important step in the direction of realizing a more reliable monitoring of the global progresses towards the goal of food security for all people worldwide (as expressed by Target $2.1$ of the Sustainable Development Goals) and, as such,  it is expected to gain increasing attention by practitioners and decision makers in the field.

Additionally, a second study investigated the issue of comparability between food insecurity scales referred to households without children (Adult scale) and  households with children (Children scale), within each national context. Results show that the national thresholds currently used to compute prevalences of food insecurity at \emph{nominally} the same level of severity among households with and without children might not always represent the same degree of the restrictions on food access. This seems especially evident for the most severe category of food insecurity, where current thresholds on the Children scale are lower than those computed as equivalent to the thresholds used on the Adult scale for this category.

Future studies are expected to shed light on possible reasons and additional aspects of this topic. A more detailed characterization of the phenomenon of food insecurity across countries as well as in households with and without children (especially from a social and economic point of view) will likely better motivate and clarify the distinctive behaviour of different food insecurity scales.
Furthermore, similar analyses to be conducted on other experiential scales of food insecurity might contribute to reach a deeper knowledge of the phenomenon. Significant examples being the HFSSM in North America, as well as applications of ELCSA in countries of the Latin America beyond the ones here considered.

%\bibliographystyle{spmpsci.bst}
%\bibliography{mybibliography.bib}

\clearpage
\subsection*{References}
\renewcommand{\bibsection}{}
\bibliography{author.arxiv}
 
\bigskip 

\bigskip

\small {\bf Authors' address:}

\bigskip

\noindent University of Rome La Sapienza, 
\hfill {\tt onori.federica@gmail.com}\\
Faculty of 
Information Engineering, Informatics and Statistics, \\
Department of Statistical Sciences, \hfill {\tt pierpaolo.brutti@uniroma1.it}\\
Italy
\bigskip 

\noindent Food and Agriculture Organization of the United Nations, \hfill {\tt sara.viviani@fao.org}\\
Statistics Division (ESS),\\
Italy

\end{document}